# Seeing Science

Alyssa A. GOODMAN[1,2]

The ability to represent scientific data and concepts visually is becoming increasingly important due to the unprecedented exponential growth of computational power during the present digital age. The data sets and simulations scientists in all fields can now create are literally thousands of times as large as those created just 20 years ago. Historically successful methods for data visualization can, and should, be applied to today's huge data sets, but new approaches, also enabled by technology, are needed as well. Increasingly, "modular craftsmanship" will be applied, as relevant functionality from the graphically and technically best tools for a job are combined as-needed, without low-level programming.

## 1. Introduction

The essential function of data visualization is to offer humans a way to see patterns in quantitative information that would otherwise be harder to find. Many people today believe that computers can always find these patterns as easily, or more easily, than people can. The people who do *not* believe computers have this power fall into two groups: researchers who strive to create tools as good as humans, and small children (who have not yet been indoctrinated to believe that computers are superior computers to humans in all ways!). The most productive research in data visualization today is focused on developing *technology to augment the human ability* to find patterns.

## 2. History

Before the introduction of the computer into science, data visualization took two forms: 1) hand-drawn sketches made by researchers themselves; and 2) professionally-drafted illustrations. Some "conventions" for making these drawings did develop (e.g. Cartesian coordinates), but the makers of early scientific drawings were free to draw upon or create whatever tools and rubrics were most appropriate to their tasks, conventional or not.

As computers entered the picture, several important changes took place. First, on the upside, the amount of data scientists could generate and analyze began to rise very rapidly, and the alternatives available for how to display it (e.g. animation, 3D graphics) began to expand. On the downside, the tools that were developed to put data visualization into the hands of the scientists themselves typically offered nowhere near the level of flexibility and craftsmanship that the combination of hand-drawing and professional draftspeople could. As a simple example, think about how easy it is for a person to write a name along a curving river or street in a map (Figure 1), but how much harder it is to get a computer to do that just as well.

Today, the very best tools available for data analysis and visualization are being developed with attention to the kinds of graphical details and functionality that the work of draftspeople used to add to science. Below, I argue that what is needed now is for high-craft tools to be made modular and interoperable enough so that scientists can combine the functionality offered by various systems into ones where "modular craftsmanship" is possible.

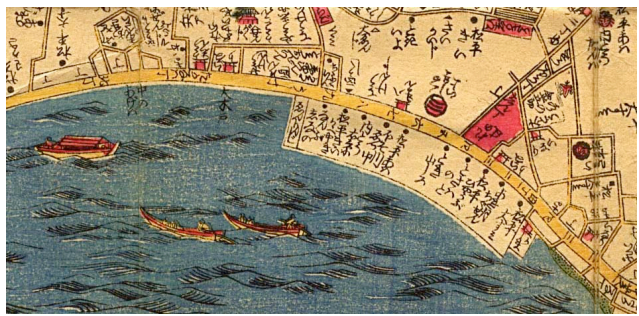

Fig.1. An historical map of Edo (1844-48). Notice the craftsperson's attention to orientation in the labeling, and the beautiful details of illustration. [1]

## 3. Data • Dimensions • Display

Formally, we can frame visualization challenges by thinking about interactions amongst *data*, *dimensions*, and *display*. Some *data* to be visualized arise from continuous functions (e.g. fitting), others come from discrete measurements (e.g. observational/experimental data). Some *data* sets are inherently large and others small. Most data sets have either an inherent *dimensionality*, or dimensionality imposed when a choice is made about what quantity/quantities are to be explored/displayed as functions of others. For example, brain imaging data is often <u>three</u>-dimensional, but is often displayed as a series of <u>two</u>-dimensional slices. Oftentimes, it is the nature of a *display* mode (e.g. monochrome vs. color, paper vs. electronic, static vs. dynamic, etc.) that sets boundaries on what *data* are *displayed* with what *dimensionality*.

The word "dimensionality" should not be taken too literally. In some cases, such as medical, geospatial or astronomical data data, there are natural coordinates, with a natural number of dimensions (typically 3 plus time) in which one can display sensed quantities. Often, that kind of "natural" display is particularly useful. But, even in fields offering what seem "natural" combinations of dimensions, there are often abstract combinations of "dimensions" (e.g. a 3D plot of global temperature vs. time and wealth) that are equally or more instructive. And, in fields of study without natural spatial dimensions (e.g. wealth management) abstract combination of variables, for analysis in multi-dimensional space, is the norm.

[1] Professor of Astronomy & Founding Director of the Initiative in Innovative Computing, Harvard University
[2] Scholar-in-Residence, WGBH Boston

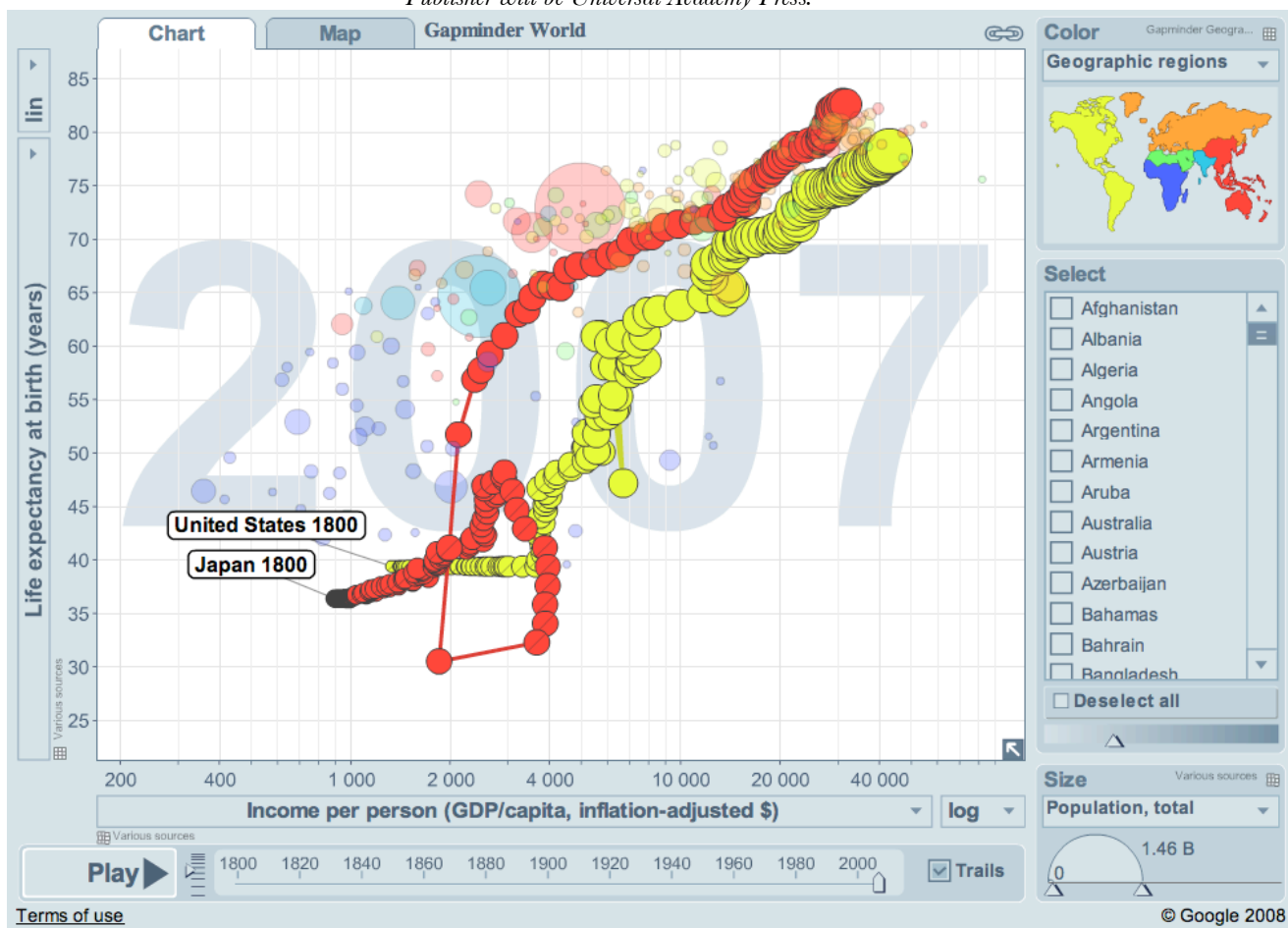

Fig. 2: A snapshot of the "Gapminder" web tool (from http://www.gapminder.org) as applied to data showing life expectancy vs. income, highlighting the U.S. and Japan. In an online version of this document, click here to try out the interactive features of this graph. The Gapminder tool was originally developed by Dr. Hans Rosling, in conjunction with the development of the Trendalyzer Software Package. Trendalyzer was acquired by Google, Inc. in March 2007.

So, then, the general challenge in "Seeing Science" is to create and use tools appropriate to the kind of *data* available, with regard to the data's either natural or potential *dimensionality*, whilst taking into account the constraints––or opportunities ––offered by particular form factors of *display*.

## 4. New Options

Some of the options open to us today for displaying our data are essentially the same as, or very closely based on, those open to the previous generation of scientists. For example, an *x-y* graph using different symbols to indicate one or more measured "*y*" quantities is still often an excellent option for exploring data or communicating its import.

Fig. 2 shows a static screenshot of a(n interactive) graph created using the gapminder.org web site [2]. The central panel of the Figure essentially is just an *x-y* graph of the relationship between life expectancy and time for the U.S. and Japan from 1800 to 2007. This graph on its own is quite informative, and for the purposes of a two-*dimensional*, static, printed, document like this one, it is a suitable *display* of these *data*.

Yet, as the many grey-backgrounded panels around the central one in Fig. 2 suggest, much more interactive investigations are possible when *display* features not typically available to previous generations are used. As users of gapminder (or, more generally, "Trendalyzer" or "Motion Chart" [5], know, graphs are generated after users interactively select which data sets to explore (notice the panel at the right in Fig. 2, listing the selectable countries.) Then, users can hit the "Play" button, to see relationships evolve over the range of time selected in the panel at the bottom, and the *animated* time series generated can be recorded (and thus summarized) as the *x-y* graph you see in Fig. 2. The size of the symbols used in fact represents a third "*dimension*" of information in these *displays*, as it is set to be proportional to a country's population.

The "`Map`" tab at the top of Fig. 2 allows users to display the same global health *data* sets that underlie all of gapminder.org in a "natural" geospatial context. Yet, the geospatial view is not always the most relevant in studies of global health, and the greatness of gapminder comes from the flexibility it offers users to explore or demonstrate relationships within and amongst *data* sets, using more than two *dimensions* at once, by making use of the options offered by a dynamic, color, two-dimensional *display*.

Keep in mind that the word "*dimensions*" here really means "variables." In that sense, Fig. 3, at right, shows what looks like a "three dimensional" view of gas in interstellar space, but in fact, only two of the dimensions are purely spatial ("R.A." and "Dec". are coordinates measured on the sky). The third dimension is "$v_z$", a velocity measure that can be used to separate objects that otherwise project on top of each other in plane-of-the-sky-only (2D) views. The paper from which Fig. 3 is drawn seeks to study relationships between the objects shown with darker shading, but even with the static 3D visualization shown here, those relationships are not sufficiently apparent. Instead, it was necessary to publish the paper *displaying* the figure in a way that allows the user to "turn" the figure around in arbitrary directions, so that expert readers of the paper can see for themselves how connected or disconnected features are from various vantage points. So, the *data* were *displayed* as an interactive "three-*dimensional*" figure (see caption) that presents as a pretty-good display on a static page, but as a much better, more informative, one on an interactive 2D computer monitor.

Both gapminder (Fig. 2) and 3D publishing (Fig. 3) offer new insights into more *dimensions* of *data*, when more aspects of *display* technology are used.

Yet, the options offered by technology can be overwhelming, which today leads both to their under- and over-use. Many researchers are presently unaware of re-usable tools like the software that underlies gapminder.org, or the 3D capabilities of Adobe Acrobat and other programs[3], which leads to under-use. Conversely, when researchers whose main talents lie in their domain specialty, and not in graphics or visualization, encounter new tools, they sometimes over-use the tools to the detriment of communicating their science. For examples of that kind of over-use, think about all those mid-to-late 1990's web pages with gratuitous spinning things, flashing lines, and crazy colors, when html first came to "everyone." Or, worse, think about bar graphs made in Excel using "three-dimensional" bars where the third dimension has no meaning at all.

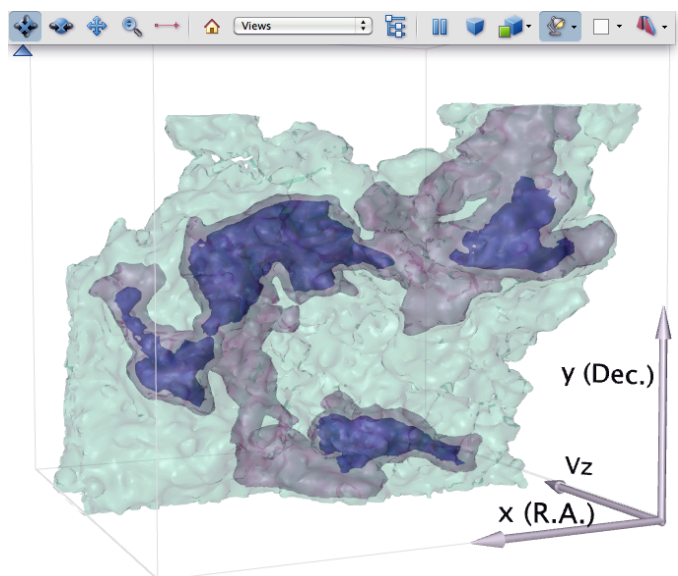

Fig. 3: A "three-dimensional" interactive figure, showing the structure of gas in a star-forming region, based on the figure published as the first "3D PDF" in the scientific journal *Nature* [3, 5].

Thus, today's scientists find themselves presented with a dazzling array of new, high-tech, tools for displaying data, but not always with enough expert guidance to use those tools to maximum advantage. Further, scientists often discover that the great new tools do "almost" exactly what they need, but not quite. So, the choice becomes: use the "same old" software used in their specialty for years; or spend large amounts of time and money (which they usually do not have) re-inventing software someone else has *almost* created for them.

## 5. Modular Craftsmanship

It is well-known that the data visualization and analysis challenges faced by nearly all fields of quantitative research are largely shared [4]. They are not, however, fully identical. Programs intended for generalized data analysis and display, most notably Microsoft Excel, are fine at some level of analysis, but are usually not an end-to-end solution. They also do not, by default, produce "optimal" graphical

---

displays or handle the special needs (and/or formats) of particular disciplines.

In other aspects of scientists' lives, they are offered opportunities to use software in modular or configurable ways. For example[4], when the same scientist who struggles to find or build the right visualization tool for her data wants to figure out which hotel to stay in for a meeting, a plethora of choices to retrieve (hotels.com), compare (kayak.com) display (Google Maps API) and archive (browser bookmarks) and organize (e.g. TripIt.com) a travel search is available. And, those services can be easily organized (iGoogle) and combined into customized, re-usable (kayak.com) searches and tools.

The same kind of modularity and interoperability should be possible when it comes to creating interactive, high-dimensional, graphical displays of data. We are beginning to see an overall move toward more agile software architectures, so it is not as hard as it once was to envision how the opportunities for "modular craftsmanship" in data visualization could become commonplace.

So many of the challenges inherent in visualizing high-dimensional data on standard displays are shared, that modules which each handle a different kind of data/dimensional combination can be envisioned. Google's creation of open Visualization APIs is clearly a step in this direction [5].

Today, astronomical imagery is often best displayed using modular tools and APIs spun off from commercially-viable software designed for viewing geospatial data and/or for gaming. Google Sky came from Google Earth/Maps and the associated Google mapping APIs [5]. Microsoft's WorldWide Telescope [5] is enabled by DirectX, a collection of APIs developed primarily for running games.

## 6. The Future

In the future, it will be necessary to train scientists and learners to "see" science using an ever-changing, but easy-to-combine set of tools. It is up to today's scientists now to work as closely with today's craftspeople (software developers) as they did with yesterday's (draftspeople) to create the tools we need to accomplish the ease and flexibility in data visualization "modular craftsmanship" should allow.

And, thanks to the increasing volume and use of quantitative information in all aspects of our lives, much of the visualization software developed for commercial purposes can be re-used within the scientific community. In cases where challenges are directly shared, as is the case with, for example geospatial and astronomical data, public-private (scientist-industry) partnerships can easily be imagined, and should be pursued.

---

[4] The methods and travel-related web sites mentioned in parentheses in this paragraph are current as of November 2009. They are chosen as representative, and citing them is not intended as an endorsement by the author.